# Dual tunability of selective reflection by light and electric field for self-organizing materials

Mateusz Mrukiewicz[1]*, Martin Cigl[2], Paweł Perkowski[1], Jakub Karcz[3], Věra Hamplová[2], Alexej Bubnov[2]

[1]*Institute of Applied Physics, Military University of Technology, Kaliskiego 2, 00-908 Warsaw, Poland*

[2] *Institute of Physics, Academy of Science of the Czech Republic, Na Slovance 2, 182 00 Prague 8, Czech Republic*

[3]*Institute of Chemistry, Military University of Technology, Kaliskiego 2, 00-908 Warsaw, Poland*

*Corresponding author: mateusz.mrukiewicz@wat.edu.pl

ORCID numbers:

Mateusz Mrukiewicz: 0000-0002-0212-4520

Martin Cigl: 0000-0002-7446-3687

Paweł Perkowski: 0000-0001-7960-2191

Jakub Karcz: 0000-0003-4783-0469

Věra Hamplová: 0000-0001-9160-2477

Alexej Bubnov: 0000-0002-6337-2210

**Abstract**

The oblique helicoidal structure is formed in right-angle cholesterics under the applied electric field. The electric field changes the pitch and cone angle but preserves the single-harmonic modulation of the refractive index. As a result, in such a supramolecular system, we can tune the selective reflection of light in a broad range. Here, we report that structural colors can be tuned by simultaneously illuminating the structure with UV light and the action of an electric field. The cholesterics with the oblique helicoidal structure were doped with newly designed rod-like, chiral, and bent-shaped azo-photosensitive materials characterized by a very low rate of thermal back cis (Z) – trans (E) isomerization. The E-Z isomerization of the photo-active compounds under UV light causes the red shift of the selective light reflection in the cholesteric mixtures. We found that the molecular structure of the photosensitive materials used affects the reflection coefficient, bandwidth, response time to UV irradiation, and tuning range. The effect was explained by considering the effect of molecular matching, cis-trans isomerization, and electric field action. We investigated the dynamics of molecular changes in the oblique helicoidal structure under the influence of external factors. The designed supramolecular system has the potential application in soft matter UV detectors.

**INTRODUCTION**

Considerable efforts of the scientific community are focused on the design and investigation of the photo-active chiral self-organized systems (i.e., those possessing the liquid crystalline behavior) [1–5] or photoinduced chirality induction. Functional self-organized systems build-up of the photo-active azobenzene compounds and chiral liquid crystalline materials, like chiral nematics, may give exception input while reaching the advanced optical properties targeted to specific smart applications. In this case, the molecular structure plays a crucial role in successfully obtaining the desired photonic systems [6,7]. From the point of view of materials science, one of the most important directions in optics is the development of materials in which we can control the wavelength of reflected or transmitted light simply by changing the external fields. Generally, the selective reflection of light is observed in chiral nematic (cholesteric) and tilted smectic liquid crystals formed by chiral rod-like molecules or in mesogenic achiral mixtures doped with chiral additive [8]. In the cholesteric (N*) phase, the molecules form a periodic helical structure where a local director $n$, which is the unit vector along the local molecular orientation, rotates around the helicoidal axis, being perpendicular to the axis everywhere, $\theta = 90°$. Planarly oriented cholesteric slab selectively reflects single circular polarization of light of the same handedness, while the opposite handedness propagates through the cell unhindered. The wavelength $\lambda_{Ch}$ of selectively reflected light is proportional to the helical pitch $P_0$, and the average refractive index $\bar{n}$ and the angle of incident light $\Theta$, which is given by the following formula: $\lambda_{Ch} = \bar{n} \cdot P_0 \cdot \cos\Theta$. This Bragg reflection occurs in the visible spectrum when $P_0$ is in the order of the wavelength of the incident light propagating along the axis of the helicoid. The helical pitch and, thus, the wavelength of reflected light cannot be tuned in standard cholesteric materials using the electric field. The reason is the relationship between elastic constants of rod-like molecules, where the bend elastic constant $K_{33}$ is larger than a twist $K_{22}$. The applied electric field causes the unwinding or distortion of the helicoidal structure [9]. Due to the lack of tunability, cholesteric materials have not been considered for application as electrically tunable color filters.

In 1968 Meyer [10] and de Gennes [11] independently predicted the phenomenon of the oblique helicoidal structure, Ch$_{OH}$. This Ch$_{OH}$ structure exists under the applied electric field in cholesteric materials in which the bend elastic constant $K_{33}$ is several times smaller than the twist elastic constant $K_{22}$ [12,13]. Such a molecular structure, representing a new type of soft matter system, could not be obtained for a long time due to the lack of materials meeting the condition: $K_{33} \ll K_{22}$. Only effective utilization of new flexible dimeric mesogens make it possible to meet the required condition. The targeted mesogenic molecule that ensure the requirements is constructed by two rigid units connected by a flexible aliphatic chain. An example of such a molecule is 1,7-bis(4-cyanobiphenyl-4'-yl)heptane (**CB7CB**) [12,13]**,** which tend to favor bend deformation due to the odd number of methylene groups in the aliphatic chain connecting two rigid, rod-like cyanobiphenyl groups. It was discovered that this new molecule exhibits the previously unknown twist-bend nematic phase, N$_{TB}$ [14]. The N$_{TB}$ phase is an example of a spontaneous formation of a chiral symmetry structure with a period as short as 8-10 nm [14–16].

By adding the chiral dopant to flexible dimers, Xiang et al. demonstrated experimentally the oblique helicoidal structure and electrically tunable selective reflection [9]. Compared to the right-angle cholesteric structure, the local molecular orientation in Ch$_{OH}$ materials is different as it is tilted ($\theta < 90°$) with respect to the helicoidal axis. The cholesteric pitch $P$ and the cone angle $\theta$ of Ch$_{OH}$ can be tuned with the electric field below a threshold value $E_{NC}$ [17]:

$$E_{NC} = \frac{2\pi K_{22}}{P_0 \sqrt{\varepsilon_0 \Delta\varepsilon K_{33}}}, \qquad (1)$$

where $P_0$ it is the pitch in the absence of the electric field, $\varepsilon_0$ is a vacuum permittivity. Dielectric anisotropy $\Delta\varepsilon$ in Eq. (1) is defined as $\Delta\varepsilon = \varepsilon_\parallel - \varepsilon_\perp$, where $\varepsilon_\parallel$ and $\varepsilon_\perp$ are dielectric permittivity values measured along and perpendicular to the director, respectively. Under a sufficiently strong electric field, the molecular director realigns while the helicoidal twisting is preserved, and the helicoidal axis remains parallel to the applied electric field $E$. According to the model presented by Xiang [17], the helicoidal pitch is inversely proportional to the applied electric field:

$$P = \frac{2\pi}{E}\sqrt{\frac{K_{33}}{\varepsilon_0 \Delta\varepsilon}}. \qquad (2)$$

When the electric field decreases below a specific value $E_{N*C}$, given by the following formula [17]:

$$E_{N^*C} \approx E_{NC}\frac{K_{33}}{K_{22}+K_{33}}\left[2 + \sqrt{2\left(1 - \frac{K_{33}}{K_{22}}\right)}\right], \qquad (3)$$

the Ch$_{OH}$ structure is reorganized to the right-angle helicoidal structure. Due to these properties, the wavelength of reflection in cholesterics with the oblique helicoidal director $\lambda_{ChOH}$ is tunable in a wide spectral area through the ultraviolet (UV), visible (VIS), and infrared (IR) regions, according to the formula:

$$\lambda_{Ch_{OH}} = \frac{\left(n_o + n_e^{eff}\right)}{2}P. \qquad (4)$$

The $\lambda_{ChOH}$ parameter is defined by the effective refractive index $n_e^{eff} = \frac{n_o n_e}{\sqrt{n_e^2 \cos^2\theta + n_o^2 \sin^2\theta}}$ [17,18], where $n_o$ and $n_e$ are ordinary and extraordinary refractive indices, respectively. The electric field is the most frequent external stimulus used to tune $\lambda_{ChOH}$. However, the structural colors of Ch$_{OH}$ can also be modified by using a magnetic field [19], boundary conditions [18], temperature [20], light [21–24], and angle of incident light [25]. Such a structure can be used to realize a highly desired electrically tunable laser [26].

So far, the methods and materials that can be applied to tune the optical properties of Ch$_{OH}$ by light are quite limited. Nava et al. demonstrated the tuning of the helical pitch in helicoidal cholesterics by changing the power of laser irradiation [21,23]. Yuan et al. presented the possibility of the helical structure transformation by the chiral dopant change between positive and negative values of the helical twisting power (HTP) [27]. Furthermore, Thapa et al. used the cholesteric mixture doped with azoxybenzene molecules to tune the helical pitch by UV light irradiation [22]. It has been shown that the change of selective reflection wavelength is caused by the increase of the bend elastic constant upon UV light irradiation. The azoxybenzene compounds used in this work [22] are characterized by a relatively slow relaxation after turning off the UV light. Chornous and co-workers [24] investigated how the concentration of the chiral azo dopant affects the twisting force and phase transition temperatures of the helicoidal cholesteric mixture under different UV conditions.

The main objective of this work is to design a smart self-organizing supramolecular photo-active system and to prove the possibility of the dual tunability of selective reflection ensured simultaneously by the irradiated light and the applied electric field.

## MATERIALS

This section describes the structures of the photo-active materials designed and used as a functional dopant as well as the design of the targeted multicomponent mixtures.

### Photo-active dopants

New photosensitive dopants **4DCN** (Figure 1a), **6DAHL** (Figure 1b), and **8BVJH12** [28] (Figure 1c) were designed and synthesized at the Institute of Physics of the Czech Academy of Sciences. The **4DCN** compound is a rod-like photo-active material composed of a laterally substituted azobenzene rigid core, with a terminal cyano group on one side and an alkyl terminal chain on the other side. Compound **6DAHL** is a chiral azo compound having lateral methyl substitution in *ortho*-positions of the azobenzene structural motif. The molecular structure of the bent-shaped compound **8BVJH12** is based on 1,3-disubstituted benzene central ring with a directly attached azo group.

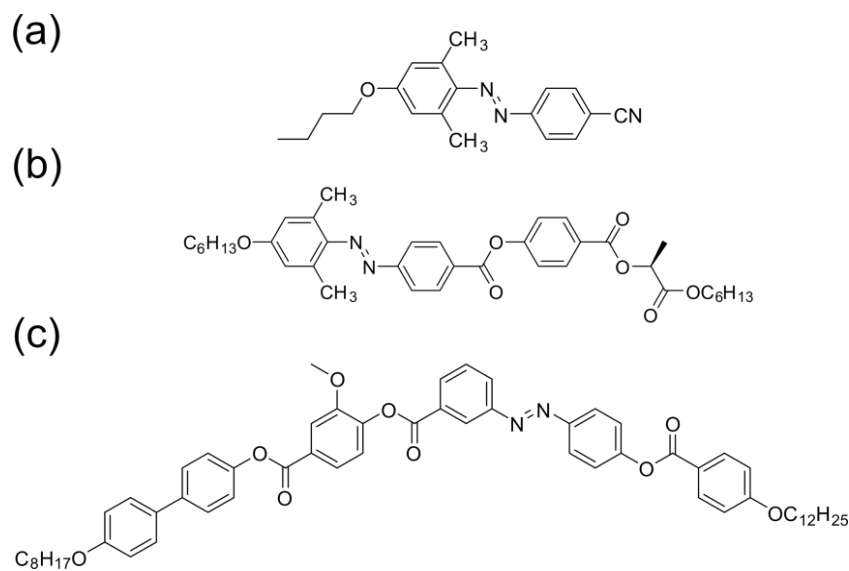

Figure 1. Molecular structure of the photo-active dopants (a) achiral **4DCN** (b) chiral **6DAHL** and (c) bent-shaped **8BVJH12**.

The design and synthesis of the photo-active dopant **BVJH12** were reported previously [28]. The two new photo-active materials, **4DCN** and **6DAHL** were synthesized as presented in the synthetic scheme in Figure 2. The achiral dopant **4DCN** was synthesized starting from 4-aminobenzonitrile **1**, which was diazotized with sodium nitrite at low temperature, and the formed diazonium salt was coupled with 3,5-dimethylphenol under basic conditions, giving azo compound **2**. In the second and final step of the synthesis, the hydroxyl group of intermediate **2** was alkylated with 1-bromobutane yielding **4DCN**. The chiral photo-active dopant **6DAHL** was synthesized from azo compound **3**, which was synthesized as described previously [6]. Starting material **3** was alkylated by 1-bromohexane, and subsequently, the ester group was hydrolyzed under basic conditions giving acid **4**. Finally, acid **4** was esterified with chiral hydroxybenzoate **5**, synthesized according to Ref. [29], in a DCC-mediated reaction to yield target chiral dopant **6DAHL**. The experimental part of the synthesis and synthesis procedures were described in more detail in Supplementary Material.

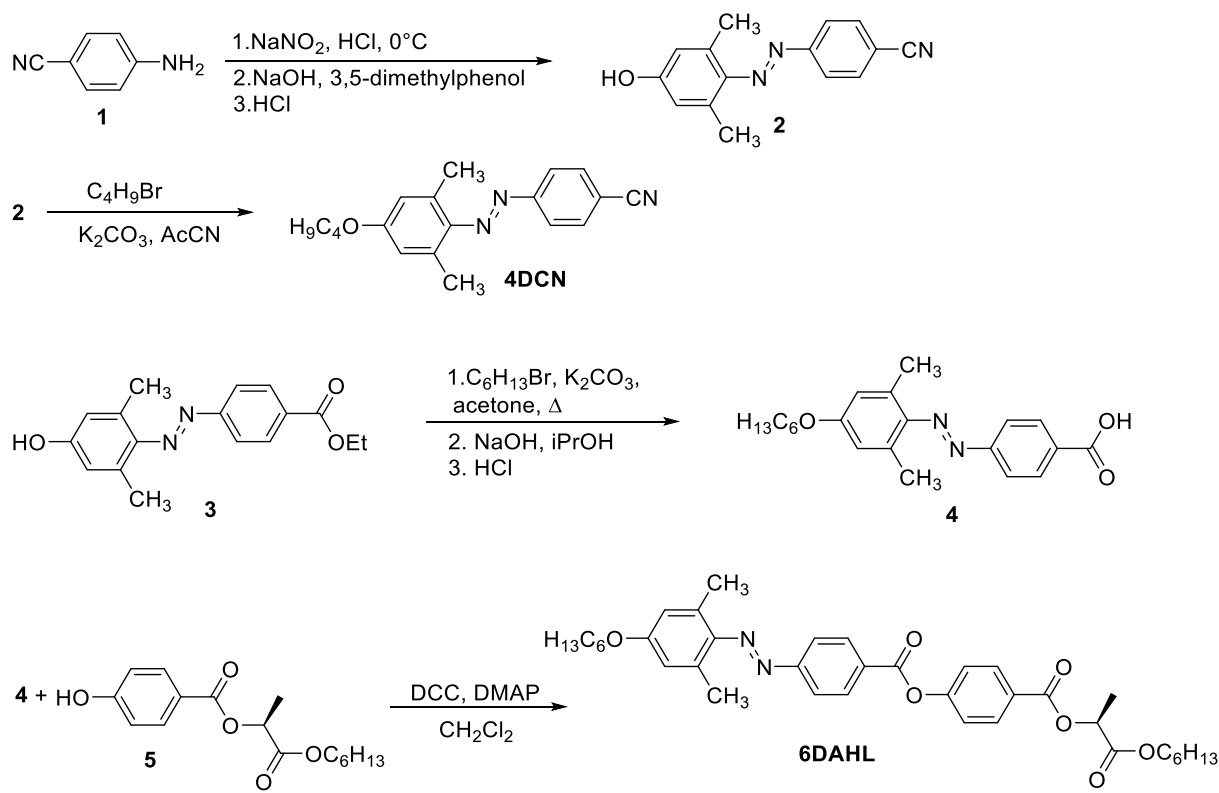

Figure 2. Synthesis of the photo-active achiral **4DCN** and chiral **6DAHL** dopants.

The molecular structure of the photo-active **4DCN**, **6DAHL**, and **8BVJH12** dopants, in the form of E-isomers and corresponding photogenerated Z-isomers, was visualized using the open-source software Avogadro (Figure 3). The molecular conformations were calculated by a DFT calculation on a B3LYP 6-31Gd level of theory. The E-isomer of the achiral **4DCN** molecule has an elongated rod-like shape, similar to that of the **5CB** (4-Cyano-4'-pentylbiphenyl) [30] molecule. The calculated length of the E-isomer is $l \cong 18$ Å, which is comparable to the length of **5CB** ($l \cong 19$ Å) [31]. The chiral **6DAHL** molecule is characterized by a length of 36 Å. The molecular length of the bent-shaped (E) photo-active **8BVJH12** material is $l \cong 58$ Å, while the bent-core liquid crystal dimer **CB7CB** is found to be $l \cong 26$ Å [12]. The overall bend angle of 144° is larger than the angle between the terminal mesogenic groups of **CB7CB** (120°) [12,32]. The Z-isomers produced under UV irradiation disrupt the liquid crystal order due to their bulky shape with the bent angle of ca. 68°. The E-Z isomerization occurring in the investigated photo-active compounds will certainly change the optical properties of the liquid crystal medium and, consequently, change the birefringence and the selective reflection wavelength. The change in molecular order in cholesteric mixtures may also be influenced by the photoinduced isothermal phase transition from the liquid crystalline state to the isotropic phase driven by the isomerization process.

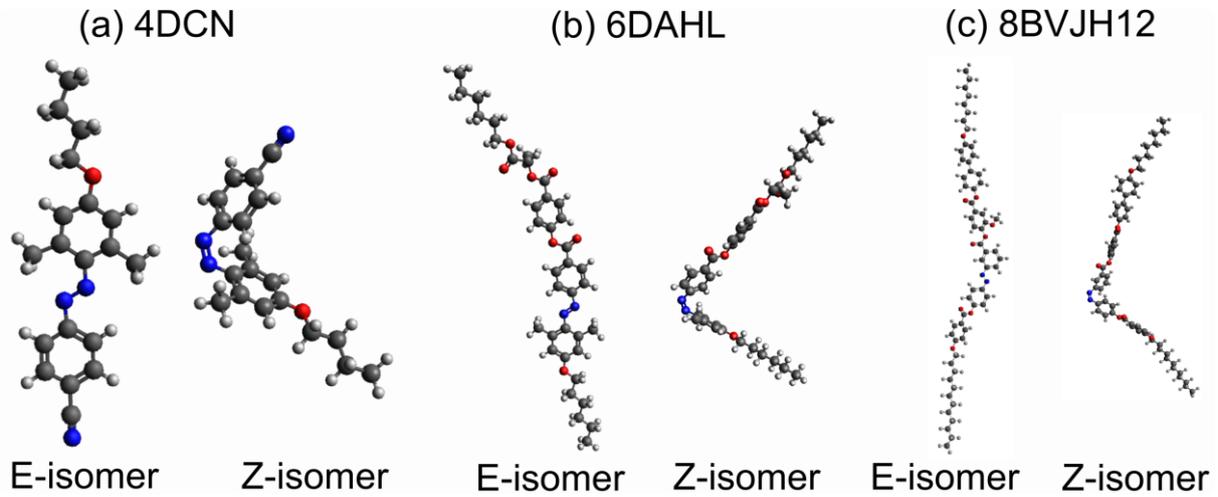

Figure 3. The molecular structure of the photosensitive dopants **4DCN** (a), **6DAHL** (b), and **8BVJH12** (c), in the form of E-isomers and the photoinduced Z-isomers.

**Design of the multicomponent mixtures**

To reach the objectives of this work, three multicomponent mixtures were designed and studied. All the materials used to create the oblique helicoidal structure, specifically **CB7CB** [12,13], and **5CB** [30] liquid crystals, and left-handed chiral dopant (S-(+)-2-octyl 4-(4-hexyloxybenzoyloxy)benzoate) [33] were synthesized at the Institute of Chemistry in the Military University of Technology, Poland.

To achieve the dual tunability of the cholesteric pitch and the wavelength of light reflection, three cholesteric mixtures with the same amount (~10% by weight) of photosensitive dopants were formulated (see Table 1). To develop the cholesteric mixture with the specific shape of oblique helicoidal director, the achiral liquid crystal **CB7CB** dimer was used. It shows a low value of bend elastic constant $K_{33}$ near the N-N$_{TB}$ transition [13]. Recent results show that the **CB7CB** dimer also provides a low value of the bend elastic constant in the multicomponent Ch$_{OH}$ mixtures, and thus, the possibility of controlling the cholesteric pitch effectively by the applied electric field [34]. The N-N$_{TB}$ phase transition occurs at a relatively high-temperature region $T_{\text{N-NTB}} = 103°C$ [13]. For this reason, the rod-like mesogenic **5CB** material was added to the cholesteric mixtures to decrease the temperature range of electrically controlled selective reflection and to provide positive dielectric anisotropy $\Delta\varepsilon > 0$. Our goal was to obtain a stable oblique helicoidal structure composed of rod-like **4DCN** molecules matching **5CB** molecules and **8BVJH12** molecules matching the shape of bent **CB7CB** and to use **6DAHL** as an alternative chiral dopant for frequently used chiral dopant **S-811**. The molecular composition of the oblique helicoidal mixtures denoted as C1, C2, and C3 is presented in Table 1.

Table 1. Molecular composition of the helicoidal cholesterics mixtures with photosensitive dopants.

| Mixture: | Molecular composition: | Weight concentration [wt%] |
|:---:|:---:|:---:|
| C1 | 4DCN/CB7CB/5CB/S-811 | 10.3/56.9/28.9/3.9 |
| C2 | 6DAHL/CB7CB/5CB | 9.4/61.3/29.3 |
| C3 | 8BVJH12/CB7CB/5CB/S-811 | 9.1/58.4/27.1/5.4 |

**EXPERIMENTAL**

For the *photo-active materials*, used as dopants, the phase transition temperatures and respective enthalpy values were determined by differential scanning calorimetry (DSC) using a Perkin–Elmer DSC8000 calorimeter (PerkinElmer, Shelton, CT, USA). The sample of ~ 5 mg, hermetically sealed in aluminum pans, was placed into the calorimeter chamber filled with nitrogen. The calorimetric measurements were performed on cooling/heating runs at a rate of 10 K/min to evaluate the phase transition temperatures precisely. The temperature and enthalpy change values were calibrated on the extrapolated onset temperatures and enthalpy changes of the melting points of water, indium, and zinc. The sequence of phases for the photo-active materials was determined by the observation of the characteristic textures and their changes in a polarizing optical microscope (POM), Nikon Eclipse E600POL (Nikon, Tokyo, Japan). The planar (homogeneous) 5 μm thick cells were filled by the materials in the isotropic phase through capillary action. The inner side of the glasses used for cells was covered by indium tin oxide optically transparent electrodes (5 × 5 mm). The heating/cooling stage Linkam LTS E350 (Linkam, Tadworth, UK) equipped with a TMS 93 temperature programmer (Linkam, Tadworth, UK) was used for the temperature control, which allows temperature stabilization within ± 0.1 K. The UV-VIS absorbance spectra were recorded using a Shimadzu UV-VIS spectrometer 1601. Spectra of a solution of 15 mg/L (23.78 µmol/L) in dichloroethane were recorded at room temperature in quartz cuvettes with an optical length of 1 cm.

The resulting *helicoidal cholesteric mixtures* were filled by capillary action in the isotropic phase and then slowly cooled (0.5 K/min) to the temperature of 3°C above the $T_{N*-NTB*}$, where the $K_{33}$ elastic constant of the bent-dimer **CB7CB** exhibits minimum value [13]. The temperature of the cell was stabilized with an accuracy of ± 0.1 K by the Linkam TMS 93 temperature controller equipped with THMSE 600 hot stage. The phase transition temperatures of the helicoidal cholesteric mixtures with different photosensitive materials were determined by the polarizing optical microscope (Carl Zeiss Jenapol). For the electro-optical measurements, we used planarly aligned cells with a thickness of 20 ± 0.1 μm. The proper planar alignment was obtained by rubbed polyimide SE-130 layers (Nissan Chemicals). The indium tin oxide (ITO) electrodes with sheet resistances of $\rho$ ~100 Ω/sq and an active area of S = 7 mm × 7 mm were applied. The cells were manufactured in the Institute of Applied Physics laboratory at the Military University of Technology, Poland.

The peaks of selective reflections were induced by the electric field of frequency 1 kHz generated by Keysight 33500B waveform generator and gained by FLC Electronics A400 high voltage linear amplifier. The electric field was applied with an accuracy of 0.01 V/μm. A small change in the electric field at constant temperature caused a shift of the selective reflection peak by several nanometers. Reflection spectra were recorded using Edmund Optics CCD VIS-NIR spectrometer. Pictures of textures were taken by Delta Optical DLT-Cam Pro digital camera in the reflection mode of the polarizing optical microscope. The values of dielectric permittivity $\varepsilon$ were determined by the impedance analyzer Agilent 4294A at the frequency of 1 kHz. In the dielectric spectroscopy measurements, the $Ch_{OH}$ structure was created by the appropriate bias field.

UV irradiation, used to tune the selective reflection of light, was generated by a Dymax Blue Wave QX4 UV lamp at $\lambda$ = 365 nm. The source of light was an LED lamp outfitted with an 8 mm diameter focusing lamp. The source can be individually programmed for intensity and exposure time. In our measurements, the power of incident light was measured by ~18 mW/cm$^2$. Light from a distance of 5 cm fell perpendicularly onto the surface of the cholesteric cell for 30 seconds.

## RESULTS AND DISCUSSION

**Photo-active dopants**

The mesomorphic properties were studied for all three photo-active compounds. It appears that the achiral **4DCN** dopant does not exhibit any mesomorphic behavior. Chiral rod-like **6DAHL** compound exhibits a broad temperature range of the chiral nematic (N*) phase. During the cooling of the planarly oriented cells, we observed the oily streaks texture typical and characteristic for the cholesteric phase [35]. The compound transforms from the isotropic phase (Iso) into N* at 80°C; the glass transition temperature is −23°C. The enthalpies of the phase transitions are −0.7 and 0.63 J/g, respectively. Broadening of the mesogenic core suppressed the lamellar phases found in the direct analog of **6DAHL** without substituents [36]. It also leads to better compatibility with liquid crystal matrices compared to similar unsubstituted materials. The bent-shaped **8BVJH12** material possesses the smectic $B_1$ phase between the isotropic and crystal phases of 20 K broad; the phase sequence obtained on cooling is Iso/150°C/$B_1$/130°C/Crystal. The mesomorphic properties are described in detail in Ref. [28]. Due to proper liquid crystalline behavior, both **6DAHL** and **8BVJH12** materials can be used as photo-active dopants in the oblique helicoidal structure with expected good miscibility with the liquid crystal host.

The absorbance profiles of **4DCN** (Figure 4a) and **6DAHL** (Figure 4b) photo-active dopants clearly show two absorption bands for the E-isomer. The first absorption band at 365 nm corresponds to the π–π* transition, while the second at around 490 nm is related to the n–π* transition. After UV irradiation ($\lambda = 365$ nm) for a specific time $t$, a noticeable decrease of the π–π* absorption band due to the E-Z photoisomerization process is observed. In the case of the **4DCN** dopant, the photogenerated Z-isomer has relatively high kinetic stability and is characterized by the slow decay of the photoinduced changes. In solution ($CDCl_3$) the photo stationary content of **6DAHL** Z-isomer (after UV irradiation) is ca. 75 % (measured by H-NMR in $CDCl_3$). An essential consequence of lateral substitution by methyl groups of **4DCN** and **6DAHL** dopants is improved kinetic stability of the Z-isomer, which can provide a significant lifetime of the photoinduced changes in the liquid crystalline material. The back thermal Z-E isomerization (relaxation) rate is relatively slow, and the back conversion to the original E-form takes ca. 15 days at 20°C and ~ 6 hours at 50°C [6]. In the case of bent-shaped **8BVJH12** dopant, the content of the UV-generated Z-isomer is ca. 95%, and the thermal Z-E isomerization takes ca. nine days at 20°C [28].

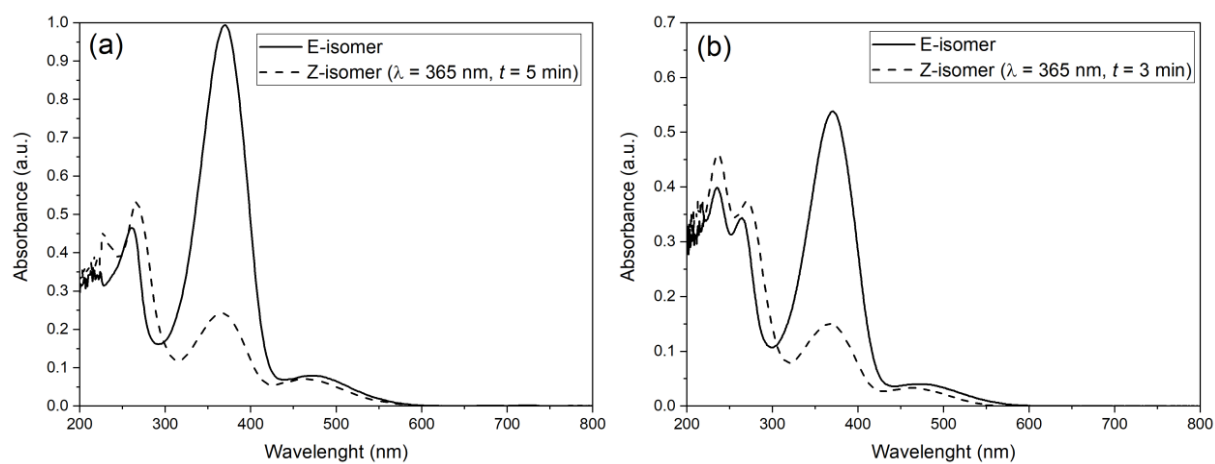

Figure 4. Absorbance profiles of photo-active (a) achiral **4DCN** and (b) chiral **6DAHL** compounds before (solid line) and after illumination by UV light at $\lambda = 365$ nm (dashed line).

**Helicoidal cholesteric multicomponent mixtures**

All the multicomponent mixtures exhibit the cholesteric phase (N*). When the temperature decreases, N* transforms at the phase transition temperature into the chiral analog of the $N_{TB}$ phase ($N_{TB}$*). The C1 mixture is characterized by the lowest $T_{N*-NTB*}$ phase transition temperature (~28°C) (see Table 2). The **4DCN** dopant has a relatively small molecule that does not tend to increase the clearing points of the mixture. In contrast, **8BVJH12** and **6DAHL** molecules, due to the presence of polar carbonyl functional groups (C=O) and thus increased intermolecular interactions, lead to higher melting points. Therefore, the highest phase transition temperatures of $T_{I-N*}$ and $T_{N*-NTB*}$ were observed for the mixture C3, 80.5°C and 53.3°C, respectively. All the mixtures are thermally stable at low temperatures, and the crystallization occurs below room temperature.

Table 2. Phase transition temperatures of studied cholesteric mixtures C1, C2, and C3 with photo-active dopants.

| Mixture: | $T_{Ntb*-Cr}$ [°C] | $T_{N*-Ntb*}$ [°C] | $T_{I-N*}$ [°C] |
|---|---|---|---|
| C1 | < 20 | 27.7 | 65.8 |
| C2 | < 20 | 44.1 | 77.7 |
| C3 | < 20 | 53.3 | 80.5 |

The transition from the right-angle cholesteric to the oblique helicoidal structure was triggered by applying the external sinusoidal electric field signals perpendicularly to the substrates in all investigated planarly aligned photosensitive cholesteric mixtures. Figures 5a-c show the certain spectral characteristics of selective reflections obtained in the absence of UV light for C1, C2, and C3 mixtures at 31°C, 47°C, and 56°C, respectively. In all cases, the intensity of the electric field ($E_B$, $E_G$, $E_O$, $E_R$) was selected to obtain the wavelength of selective light reflection corresponding to the specific colors: blue ($\lambda_B \approx 480$ nm), green ($\lambda_G \approx 550$ nm), orange ($\lambda_O \approx 600$ nm), and red ($\lambda_R \approx 680$ nm). The investigated photosensitive mixtures show a similar dependence of $\lambda_{ChOH}$ on the electric field and temperature as in the case of undoped materials, according to Refs. [17,20,26]. The values of the electric fields for which we observe a change in the selective reflection of light in the visible band are the lowest for the C1 mixture and the highest for the C3 mixture. The reason for this is the higher $T_{I-N^*}$ and $T_{N^*-NTB^*}$ phase transition temperature of the C3 mixture in comparison to that of the C1 and C2 mixtures, and therefore the increase of $K_{33}/\Delta\varepsilon$, according to Eq. (2). The increase of the key material parameter for the existence of the oblique helicoidal structure $K_{33}/\Delta\varepsilon$ with the temperature leads directly to an increase in the difference between the value of the electric field inducing the blue ($E_B$) and red ($E_R$) wavelengths. For the C3 mixture, the difference is $E_B - E_R = 0.62$ V/μm, while for C1 and C2 mixtures, it equals 0.43 V/μm and 0.50 V/μm, respectively.

Due to the consequence of the polarization selectivity rule of cholesteric liquid crystals, the reflection coefficient is at most 50% for unpolarized incident light. In the investigated $Ch_{OH}$ mixtures, the decrease in reflection below the maximum value for different wavelengths and, consequently, different electric fields can be associated with various factors. In the case of mixtures C2 and C3, a decrease in reflection coefficients for shorter wavelengths was observed (Fig. 5b, c) due to the reduction of effective birefringence $\Delta n_{eff}$ as a result of applying large electric fields to the $Ch_{OH}$ structure. According to the formula [17]:

$$\sin^2\theta = \frac{K_{33}}{K_{22}-K_{33}}\left(\frac{E_{NC}}{E}-1\right), \qquad (5)$$

the cone angle $\theta$ is smaller for high electric fields $E$, which indicates a weaker modulation of the effective birefringence $\Delta n_{eff} = n_e^{eff} - n_o$ along the helicoidal axis [20]:

$$\Delta n_{eff} = n_0\left[\left(1-\left(1-\frac{n_0^2}{n_e^2}\right)\sin^2\theta\right)^{-\frac{1}{2}}-1\right]. \qquad (6)$$

The situation is different for the C1 mixture, where the highest reflection is observed for green and orange wavelengths. In this case, the decrease of the reflection coefficient in blue is also caused by the small value of $\Delta n_{eff}$, while the red part of the spectrum is reduced by the decrease of cholesteric layers at the fixed cell thickness. Moreover, the higher electric field also narrows the spectral range for the same reason, according to Eq. (5).

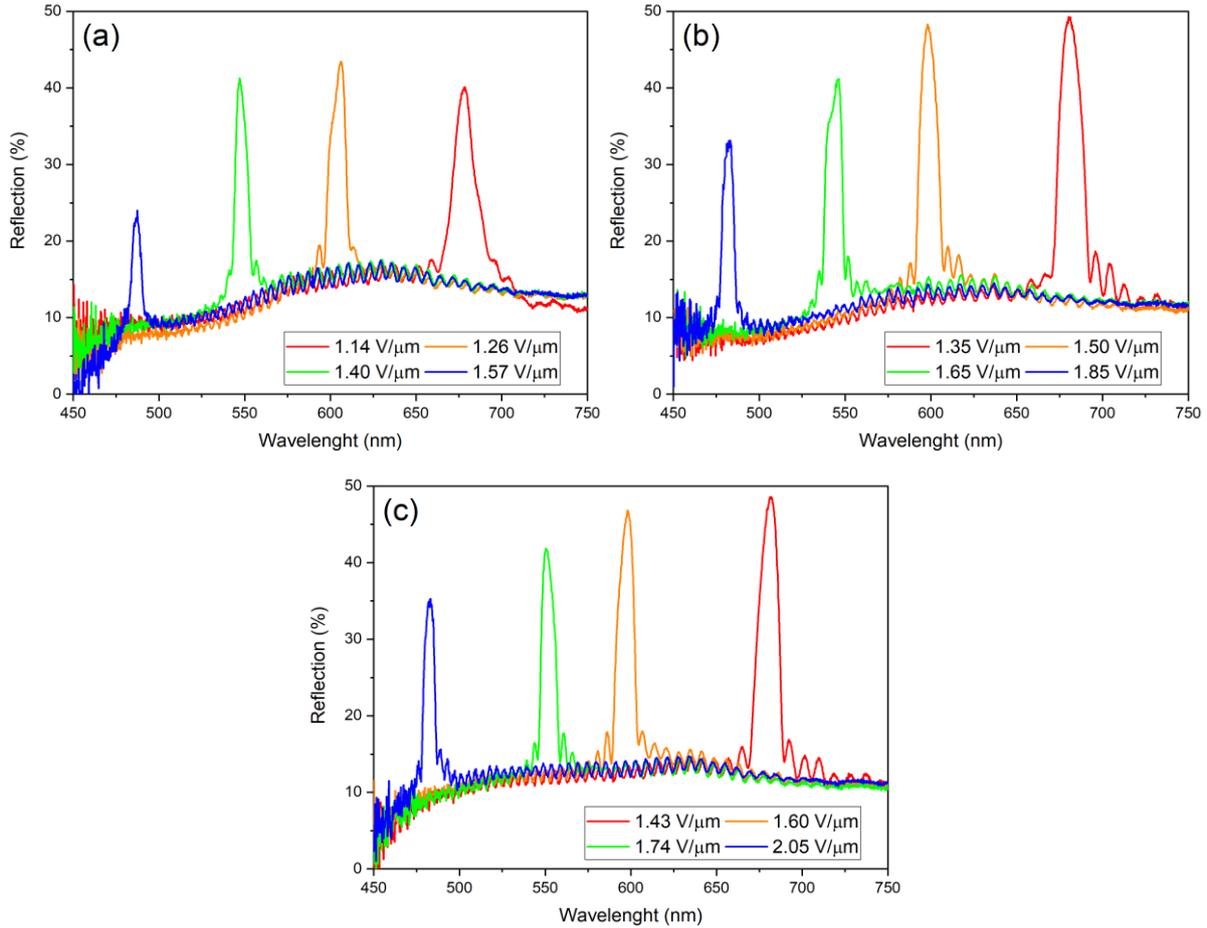

Figure 5. Spectra of electrically tunable selective reflections of red, green, orange, and blue colors obtained for the helicoidal cholesteric mixtures: (a) C1 at 31°C, (b) C2 at 47°C, and (c) C3 at 56°C with photo-active dopants (without the use of UV irradiation).

Another reason for the decline in the reflection is the limited homogeneity of structure throughout the whole sample. Figure 6 shows the electric field-controlled structural colors of the oblique helicoidal C1-C3 mixtures without the use of UV light. The microphotographs of textures were taken after the specified equilibration time (5 minutes) after the electric field was turned on. The total reflection in the C1 mixture is lowered due to the existence of slowly propagated defects (Figure 6a), which lead to the creation of $Ch_{OH}$ layers with a different pitch. These defects can be caused by disturbances in the regular oblique helicoidal structure. They arise due to a molecular mismatch of the photochromic **4DCN** molecules to the original shape of Ch$_{OH}$ formed by achiral flexible dimers, rod-like molecules, and the chiral dopant. The lower compatibility is probably due to the compound **4DCN,** which does not exhibit mesogenic behavior. Moreover, the relatively low temperature of the measurements (i.e., higher viscosity) slows down the defect annihilation process. This molecular imperfection affects the electric tunability range. On the spectrum, it corresponds to wide-band and low-intensity reflections. For this reason, we cannot observe blue reflection above 30%. The spatial inhomogeneity of the chiral structure and molecular imperfections are confirmed by the significant background measured in the plots, approximately 10%.

The cholesteric system, composed exclusively of mesogenic compounds, adapts more effectively to the electric field than non-mesogenic compounds. We noticed no significant deviations from the regular periodicity after the specific equilibrium time in the case of chiral **6DAHL** and bent-shaped **8BVJH12** dopants. Considering the textural colors, the best-ordered helicoidal structure seems to exist in the C3 mixture. The reason is the similar shape of **8BVJH12** and **CB7CB**, which is an essential material for formulating Ch$_{OH}$ mixtures. It is worth underlining that in all cases, the electric response of the mixtures was fully reversible, as after removing the electric field, the material reorganized to the original planar alignment. Even using 10 wt.% of photosensitive dopants in the resulting mixtures does not destroy the helicoidal structure.

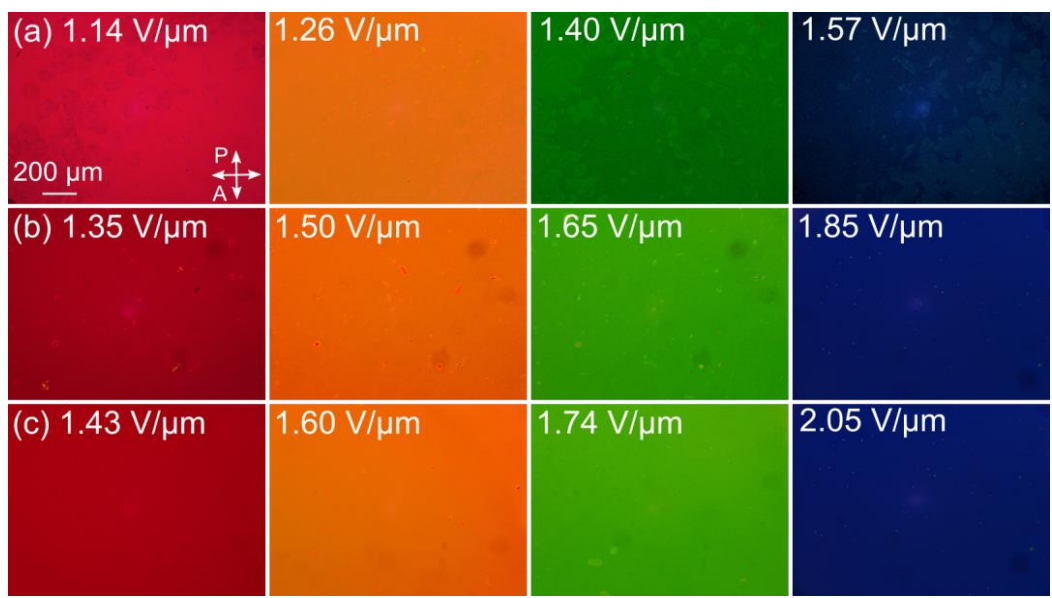

Figure 6. Structural colors observed in the helicoidal cholesteric mixtures: (a) C1, (b) C2, and (c) C3, induced by applying the specific electric fields (without UV radiation).

Electro-optical studies were also carried out using UV light to determine the phototunability of the multicomponent mixtures. The electric field was applied to C1 ($E = 1.57$ V/μm), C2 ($E = 1.85$ V/μm), and C3 ($E = 2.05$ V/μm) to induce the selective reflection at the same wavelength ($\lambda_{ChOH} \approx 480$ nm). The samples were then irradiated by UV light. After the irradiation, the electrically induced peak at blue wavelength disappears in all investigated mixtures despite the applied electric field. Irradiation of the Ch$_{OH}$ structure by UV light causes the selective reflection to shift towards longer wavelengths due to the Z-E isomerization of the photo-active dopants. However, the position of the light-induced peaks, spectral selectivity, and Z-E relaxation times significantly differ for the studied mixtures.

Figure 7 presents the shift of $\lambda_{ChOH}$ for specific time intervals. The time intervals were selected to show, whenever possible, a single reflection peak at a specific bandwidth for different parts of the light spectrum. In the case of the C1 helicoidal mixture, due to the deviation from the regular helicoidal periodicity observed for the non-UV-treated mixture, the light-induced peaks are wide-band and characterized by a low-intensity reflection (Figure 7a). The UV exposure increases the number of defects, which causes the appearance of cholesteric layers with different pitches (Figure 8a). This effect causes an additional factor lowering the reflection coefficient of the new UV-generated light peaks. The best dual tunability of selective reflection was observed for the C2 mixture (Figure 7b). Under the applied electric field, the UV-generated peaks smoothly change their position and shape in time. The structural colors seem to be much less defective than is observed in the case of C1 mixture build-up with the rod-like non-mesogenic photo-active dopant (Figure 8b). The induced selective reflection peaks in red light wavelengths appear much faster for the C2 mixture than for the C1 mixture. Specifically, for the C2 mixture, the first red-shifted peak appears at $\lambda_{ChOH} \approx 665$ nm, 5 minutes after turning off UV light, while for C1 mixture, the first peak at $\lambda_{ChOH} \approx 682$ nm is induced in 10 minutes. The rate of return to the original blue light reflection was slower for the mixture with chiral photo-active dopant. A full recovery to the initial wavelength of the selective reflection generated by the electric field takes up to several days. The photoisomerization process looks different in the oblique helicoidal mixture doped with the bent-shaped photo-active material (Figure 7c). Upon UV irradiation, the selective reflection peak is shifted only to the wavelengths of the green light ($\lambda_{ChOH} \approx 518$ nm), which was confirmed by the polarizing optical microscopy textures observed in the reflection mode (Figure 8c).

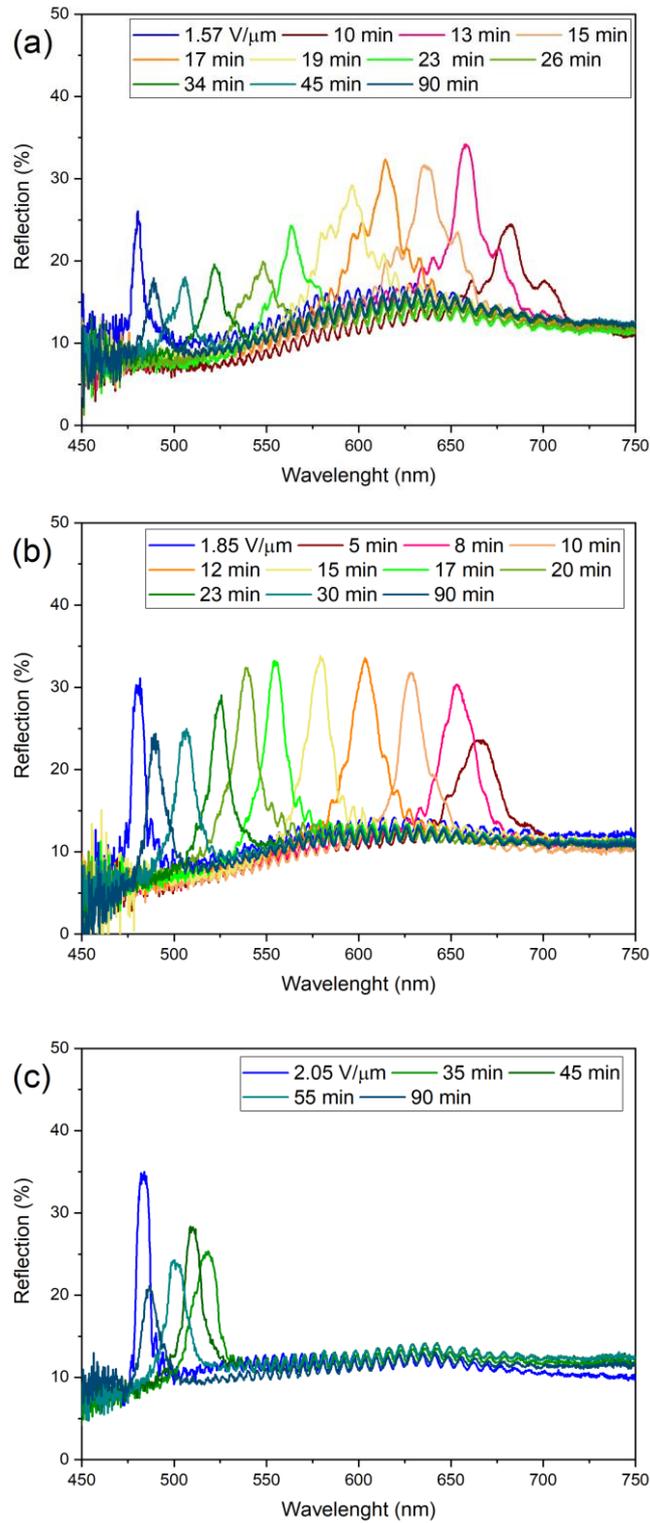

Figure 7. The shift of the selective light reflection over time under UV light irradiation in the three oblique helicoidal mixtures: (a) C1, (b) C2, and (c) C3. Initial blue reflection of light was induced by the electric fields: (a) $E = 1.57$ V/μm, (b) $E = 1.85$ V/μm, and (c) $E = 2.05$ V/μm.

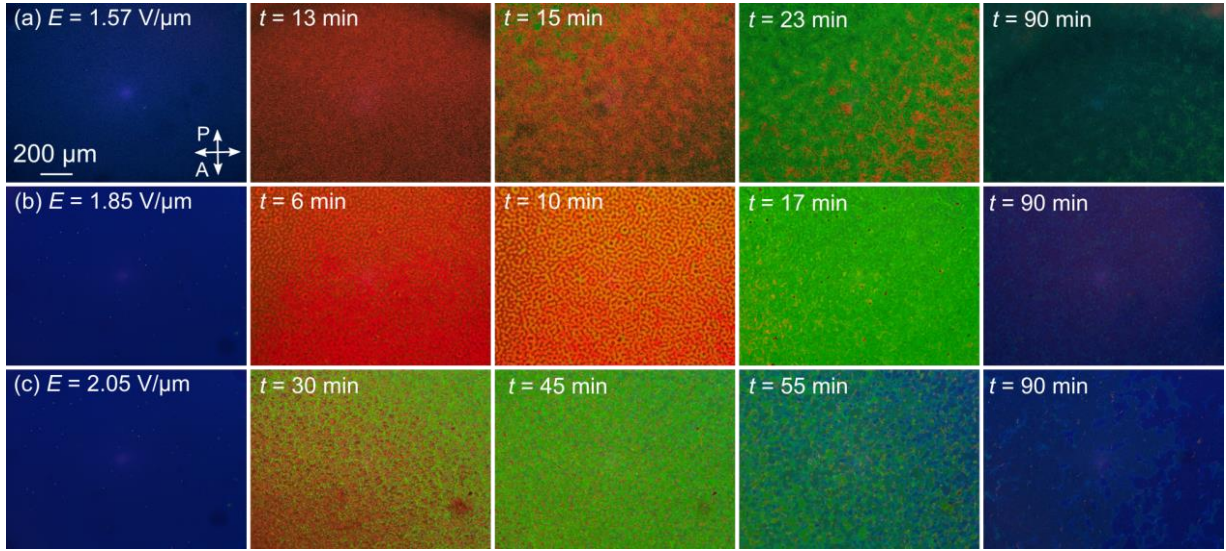

Figure 8 The structural colors observed after irradiation by UV light for (a) C1, (b) C2, and (c) C3 mixtures. Parameter *t* denotes the time after switching off UV light.

The E-Z isomerization process of individual photo-active dopants in the Ch$_{OH}$ mixtures was studied using dielectric spectroscopy measurements. At the beginning of the experiment, the dielectric tensor components along the helicoidal axis of the right-angle cholesteric ($\varepsilon_\perp$) and the unwound (by DC field) structure ($\varepsilon_\parallel$) were determined. Since the mixtures consist mainly of CB7CB and 5CB, the difference in the $\varepsilon_\perp$ values is negligible. A similar effect was observed for $\varepsilon_\parallel$. Therefore, Figure 9 shows the average values of these parameters (horizontal dashed lines). Next, the dielectric permittivity in the oblique helicoidal structure $\varepsilon_{ChOH}$ along the helicoidal axis was measured at a constant frequency. The Ch$_{OH}$ structure in the planarly aligned cells was induced by the electric field generated by the impedance analyzer. The measurements were done in the time domain, before and after UV irradiation, to investigate the dynamic of molecular changes.

After turning off UV irradiation, we observed a decrease of $\varepsilon_{ChOH}$ in all investigated cholesterics. However, the nature of the change was different in each mixture. In the case of C1 and C2, the sharp decrease in the first minute after irradiation is related to the distortion of the oblique helicoidal structure. For a short time, the liquid crystal medium becomes isotropic. Next, the electric field reshapes the helicoidal structure with Z-isomers induced by UV light, which causes the exponential increase of $\varepsilon_{ChOH}$, Figure 9. This process occurs 5 minutes after UV irradiation for C2 and a few minutes later for C1, corresponding to the first selective reflections appearing at longer wavelengths (Figures 7a and b). Two different effects are responsible for the red-shifted peaks in C1 and C2 mixtures. In the C1 mixture, UV light unwound the helicoid due to the E-Z isomerization. It thus increased the pitch length, which is observed as a more significant increase in dielectric permittivity than C2. As the cholesteric pitch increases in the Ch$_{OH}$ state, $\varepsilon_{ChOH}$ also increases towards $\varepsilon_{\parallel}$. The change of $P$ in C2 is related to the reduction of the helical twisting power. The irradiation causes a decrease in the number of chiral E-isomers and an increase in the number of chiral Z-isomers with lower twisting power. Therefore, the UV-affected oblique helicoidal structure is less twisted than the untreated one.

The C3 mixture is formed mainly of molecules (**8BVJH12** and **CB7CB**) with low bend elasticity, which prevents the complete unwinding of the helicoidal structure caused by UV light. The irradiation causes a continuous increase in the cholesteric pitch due to the photoisomerization. Therefore, the light-shifted peak does not appear from the untwisted helicoid, as observed for C1 and C2. The exponential decrease of $\varepsilon_{ChOH}$ (by only 0.3) is caused by the appearance of the light-induced Z-isomers, which slightly reduce the interaction of the medium with the electric field. The small change in the dielectric response results in the emergence of a peak in the green wavelength spectrum.

Generally, the UV-induced peak does not return the initial pre-UV blue reflection [22,37]. In the performed experiment, 2 hours after UV irradiation was turned off, the induced peak was about 5 nm higher than the initial peak. Similar behavior was observed in the dynamic measurements of $\varepsilon_{ChOH}$. The dielectric permittivity does not return to its initial value due to the long Z-E relaxation time of the photo-active dopants. The effect of the change in selective reflection under both UV light and electric field should be attributed to the difference in the bend elastic constant value reported by Thapa et al. [22] and the influence of photo-active dopants on the Ch$_{OH}$ structure presented here.

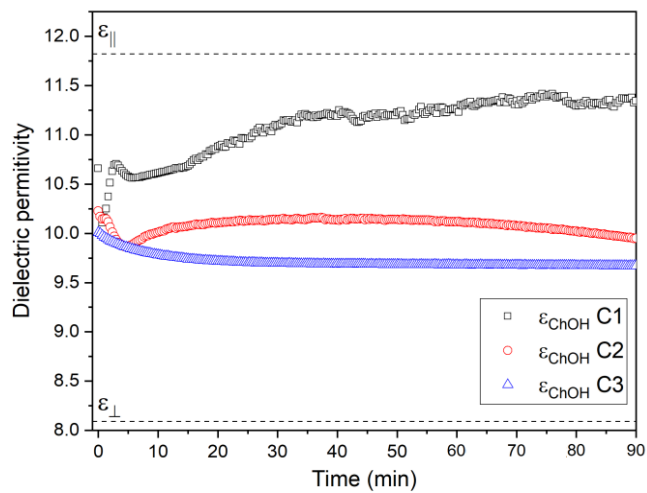

Figure 9. Dynamics of changes in the dielectric permittivity of the oblique helicoidal structure $\varepsilon_{ChOH}$ after UV irradiation.

**Conclusions**

New phototunable oblique helicoidal mesogenic system, consisting of flexible dimers and specifically designed photo-active non-chiral rod-like, chiral rod-like, bent-shaped dopants, simultaneously tunable by UV irradiation and applied electric field was designed and investigated. From a materials science perspective, we showed how to create a cholesteric liquid crystal that can be tunable in two ways. The creation of the selective reflection peak generated by UV irradiation differs considerably while changing the functional photo-active dopant structure. Materials that can be effectively used for the design of light-sensitive $Ch_{OH}$ structures should exhibit self-organizing behavior. The UV-active chiral dopant (case of C2 mixture) is the best candidate to generate the clear selective reflection shift. By carefully designing the bent-shaped photo-active molecule, it is possible to shift effectively (by UV irradiation) the selective reflection peak at a specific wavelength. The process of recovering the oblique helicoid can be described as order (before UV-illumination) – disorder (after starting UV-illumination and shortly after it) – order (the recovering time after illumination). Depending on the structure of the used photo-active dopants, this process is characterized by different time scales and effectivity. The presented results are of high importance and are essential from the point of view of $Ch_{OH}$ materials with a short response time on the applied electric field. The demonstrated smart and soft supramolecular systems can be potentially effectively utilized as UV detectors with tailored photo-response.

## Data availability

Data for this article are available in Zenodo at https://doi.org/10.5281/zenodo.14945289.

## Acknowledgment

This work was supported by NCN project 2022/06/X/ST5/01316, the bilateral Polish-Czech project MEYS 8J20PL008, NAWA PPN/BCz/2019/1/00068/U/00001, the Czech Science Foundation project 22-16499S and University project UGB 22-723. We thank Eva Otón for help with measurements using the UV lamp.

## Author contribution

MM conceived the project, performed microscopy and electro-optical studies, analyzed the data, and prepared the manuscript with inputs from all co-authors. MC, AB, and VH synthesized and characterized the photosensitive compounds. MC also performed molecular modelling. JK synthesized CB7CB. PP analyzed the data. All authors contributed to scientific discussions.